\documentclass[11pt,twoside]{article}

\usepackage{asp2004}
\usepackage{epsf}
\usepackage{psfig}
\usepackage{lscape}
\usepackage{graphicx}

\begin{document}
\title{Hubble Space Telescope Observations of the Light Echo around
V838 Monocerotis }   
\author{Howard E. Bond} 
\affil{Space Telescope Science Institute, 3700 San Martin Drive, Baltimore, MD
21218, USA}

\def\Afsar{Af\c{s}ar}
\def\HST{{\it HST}}

\hyphenation{Carl-qvist}

\begin{abstract} 
The outburst of V838 Monocerotis in early 2002, and the subsequent appearance of
its light echoes, occurred just before the installation of the Advanced Camera
for Surveys into the {\it Hubble Space Telescope\/} (\HST\/)\null. This
fortunate sequence of events has allowed us to obtain spectacular \HST\/ images
of the echoes, yielding not only pictures of extraordinary beauty, but also
providing unique scientific information.

Our team has used the \HST\/ images to provide a direct geometrical distance to
V838~Mon, based on polarimetric imaging, and limits on the distance based on the
apparent angular expansion rates.

Several morphological features seen in the \HST\/ images strongly suggest that
the illuminated dust was ejected from the star in a previous outburst, similar
to the current one. In particular, a ``double-helix'' feature points exactly
back to the star. Moreover, three-dimensional mapping of the outer edges of the
dust suggests an overall ellipsoidal shape, centered on V838~Mon itself. And the
appearance of the light echo in the most recent {\it Hubble\/} images is
remarkably similar to that of a well-known planetary nebula, M27.

Future work on the \HST\/ images will include an analysis of interstellar-dust
physics, in a situation where the scattering angle and illumination are
unambiguously known, and visualization of a fully three-dimensional map of the
dust distribution.

\end{abstract}

\section{The Hubble Space Telescope}

The {\it Hubble Space Telescope\/} ({\HST\/}) has recently celebrated its
16th anniversary in orbit, having been launched aboard the NASA space shuttle in
April 1990. Since then there have been four astronaut servicing missions between
1993 and 2002 (SM 1, 2, 3a, and 3b). We are currently hoping that there will be
an SM4 in 2008, during which two new instruments would be installed: the Cosmic
Origins Spectrograph (COS) and the Wide Field Camera 3 (WFC3). WFC3 will be a
remarkable instrument that provides direct imaging at UV, optical, and near-IR
wavelengths.

The \HST\/ is uniquely suited for high-resolution imaging. In this paper I
would like to present some results from {\it HST\/} imaging of the remarkable
light echoes around V838 Monocerotis.

\section{HST Observations of V838 Mon}

Shortly after V838~Mon reached maximum light in February 2002, an expanding
light echo was discovered by Henden, Munari, \& Schwartz (2002). These light
echoes have evolved to become  the most spectacular display of the phenomenon in
astronomical history. They have been the subject of extensive imaging by
ground-based observers (e.g., Crause et al.\  2005 and references therein), and
are of course the inspiration for the conference here in La~Palma.

Based on the highly structured appearance of the initial ground-based images,
our team proposed for Director's Discretionary (DD) time on \HST\/ for a program
of direct imaging and imaging polarimetry. The team members are as follows:
S.~Starrfield (Arizona State University); Z.~Levay, N.~Panagia, W.~Sparks,
B.~Sugerman, R.~White, and myself (STScI); A.~Henden (AAVSO); M.~Wagner
(University of Arizona); R.~Corradi (Issac Newton Group); U.~Munari (Padova
University); L.~Crause (SAAO); and M.~Dopita (ANU)\null.

We received \HST\/ observing time at five epochs in 2002 through DD allocations:
April, May, September, October, and December. All of the observations were made
with the Advanced Camera for Surveys (ACS), which had been installed in \HST\/
during SM3b in March 2002. I need not emphasize to this audience how
extraordinarily unfortunate it is that no \HST\/ observations were obtained
during 2003---the loss of this opportunity is truly incalculable. However, the
echoes were imaged twice in 2004 through the Hubble Heritage program, in
February and October. More happily, the \HST\/ Cycle~14 allocation committee did
award our team observing time for an intensive \HST\/ imaging campaign from
October 2005 to January 2006, and we also have two more epochs of observations
scheduled in Cycle~15 for late 2006 and early 2007.

A summary of results from the 2002 imaging is given by Bond et~al.\ (2003). A
discussion of the imaging polarimetry is presented elsewhere in these conference
proceedings (Sparks et al.\ 2006).

It is nearly impossible to represent the extraordinary \HST\/ images on paper,
especially in black and white. As a partial guide to what is available, Figure~1
presents a montage of the \HST\/ images from 2002 and 2004, all at the same
angular scale. However, I urge readers to go to the STScI website
(www.stsci.edu), and of course to the Hubble Heritage site (heritage.stsci.edu),
where many full-color images of the light echoes can be viewed. 

\begin{figure}[ht]
\begin{center}
\includegraphics[width=\textwidth]{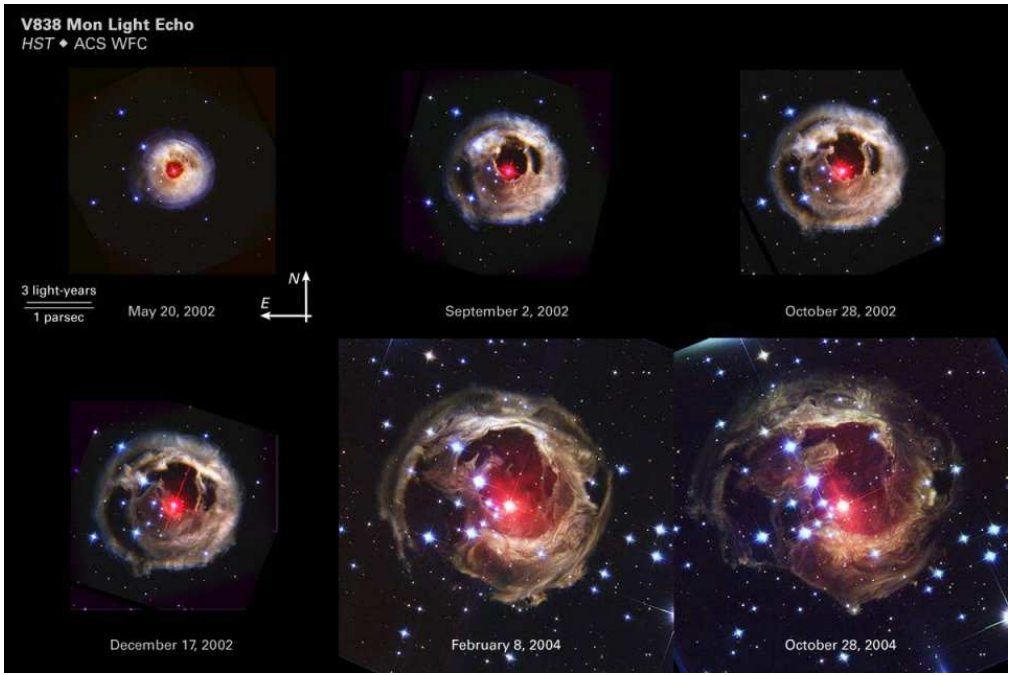}
\end{center}
\caption{\HST\/ images of V838 Mon, 2002-2004. The images are all at the same
angular scale, and dramatically illustrate the apparent expansion of the echoes.
Images taken in late 2005 and early 2006, however, have nearly the same size as
the October 2004 image, indicating that we are about to enter the phase of
apparent contraction.}
\end{figure}

An animation showing all of the \HST\/ images has been prepared by Zolt Levay
and myself. There is a link to this animation at the conference website,
www.ing.iac.es/conferences/v838mon, or you can view it at
www.stsci.edu\slash$^{\scriptscriptstyle\sim}$bond/v838mon\_2002-5\_anim.gif.

\section{Applications of the HST Images}

In addition to their aesthetic beauty, the \HST\/ images of the V838~Mon light
echoes provide several scientific applications. These include direct geometrical
distance determinations, and information on the detailed morphology and
three-dimensional structure of the circumstellar dust. (The images will also
ultimately provide important information on the physics of interstellar dust
particles, since we are in an unique situation where both the spectrum and time
dependence of the illumination, and the scattering angle, are unambiguously
known.)

\subsection{Geometrical Distance Determinations}

Light echoes provide two independent means for geometrical distance
determinations. One method is based on apparent angular expansion rates, and is
described in detail by Bond et~al.\ (2003). This method requires an assumption
about the geometry of the dust. For the 2003 analysis, we assumed thin,
spherical sheets centered on the star, and were able to obtain a lower limit to
the distance of $\sim$2~kpc.  

We will carry out a similar detailed analysis of the \HST\/ images from the
2005-6 campaign, but an approximate calculation goes as follows: In late 2005,
the radius at which zero apparent expansion would be seen (for spherical shells
centered on the star) would be $\sim$1400~light-days. The corresponding angular
radius of zero expansion would be $40''\times(6\,{\rm kpc}/d)$, where $d$ is the
distance. The apparent expansion in the late 2005 \HST\/ images is indeed close
to zero at a radius of about $40''$, consistent with an approximate distance of
$\sim$6~kpc (see next paragraph). At larger radii, the echoes are clearly still
expanding, indicating that the distance cannot be much less than $\sim$6~kpc.

The other geometrical method relies on polarimetry. This method was worked out
in detail more than a decade ago by our team member Bill Sparks (Sparks 2004).
His intention was to use the technique to find distances to extragalactic
supernovae, but the method now finds its application to a peculiar object in our
own Milky Way! The Sparks technique is based on the fact that maximum linear
polarization occurs for scattering off of dust particles at an angle of
$90^\circ$. Given the geometry of light echoes (e.g., Bond et~al.\ 2003;
Sugerman 2003), this location corresponds to a linear radius of $c\Delta t$,
where $\Delta t$ is the time since the outburst. The corresponding angular
radius of maximum linear polarization then yields the distance. As described
elsewhere in these proceedings, the \HST\/ polarimetric images yield a distance
of 5.9~kpc (Sparks et al.\ 2006). Reassuringly, this agrees very well
with the distance of 6.2~kpc recently obtained from spectral classification and
photometry of three B-type stars in the sparse, young cluster surrounding
V838~Mon (Bond \& \Afsar\ 2006; \Afsar\ \& Bond 2006).

\subsection{Morphology of the Light Echoes: Is the Dust Interstellar or Ejected
from the Star?}

In this subsection, I want to discuss the issue whether the dust illuminated in
the light echoes arises from material ejected from V838~Mon in previous
outbursts, or is merely ambient dust with no special relation to V838~Mon (as
advocated, for example, by Tylenda 2004 and Crause et al.\ 2005). My approach is
not based on physics, but upon the morphological approach described so
brilliantly by Zwicky (1957).  I conclude that the dust was indeed ejected from
V838~Mon. Apart from the fact that the star is copiously producing dust during
its current outburst, and that that dust will expand slowly away from the star
and would be illuminated by any future outburst, there are the following
morphological arguments:

\begin{enumerate}

\item {\it There are features in the dust that show a connection with the star.}
The most dramatic one is the ``double-helix'' structure seen in the \HST\/
images of September, October, and December 2002 (which was discussed by
Carlqvist 2005). The axis of the helix points directly at the star in all three
images, one of which is shown in Figure~2. It would be remarkable to find such a
precise alignment if the dust were merely ambient interstellar material that
happens to lie in the vicinity of V838~Mon.

\begin{figure}[ht]
\begin{center}
\includegraphics[width=2.25in]{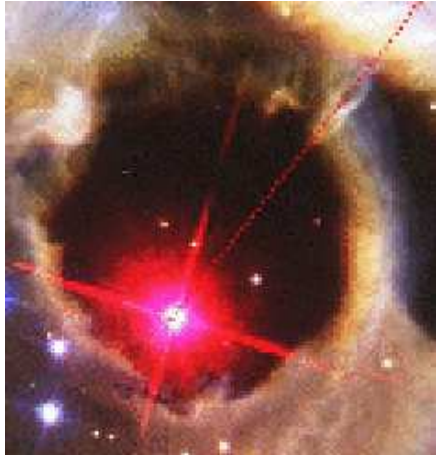}
\end{center}
\caption{Detail from  \HST\/ image of the V838 Mon light echo on 2002
September~2. A dotted line drawn from V838~Mon through the ``double-helix''
feature at the upper right passes precisely through the axis of the helix.}
\end{figure}

Note that the three-dimensional location of any illuminated dust particle is
unambiguously known in a light echo.  We plan to develop three-dimensional
visualizations of the dust distribution, which will doubtless provide further
information on the relationships between the dust and the star. At a given
epoch, the illuminated dust lies along a paraboloid with the star at its focus.

\item {\it The dust has a well-defined outer edge, and the distribution of outer
edges is centered near the star.} As a simple example of three-dimensional
mapping, I marked the outer edges of the dust distribution in each of the
available \HST\/ images. As noted in the previous paragraph, the $(x,y)$
position of any feature in the echoes at a time $\Delta t$ corresponds exactly
to a $z$ location in the line of sight ($z$ does, of course, depend also on the
distance to the star, but as noted above $d$ is now well known). 

The images in 2002 have very sharp outer edges (see Figure~1), indicating a
sharp drop in the dust density at the corresponding locations in
three-dimensional space (since the edges are much sharper than would be produced
by purely geometrical effects). The color images obtained by \HST\/ are
particularly useful, because they contain obvious sharp blue rims, corresponding
to the sharp blue peak at maximum in the light curve. The edges are not quite as
sharp at later epochs, but are still easily located. Until we can present a
fully three-dimensional visualization of the dust, I have simply measured the
locations of the outermost edges along a north-south axis passing through the
star, and along an east-west axis. I then converted these locations in the plane
of the sky to the true three-dimensional locations. The resulting maps of the
outer edges are shown in Figure~3a (north-south), and Figure~3b (east-west).

\begin{figure}[htb]
\begin{center}
\includegraphics[width=2.45in]{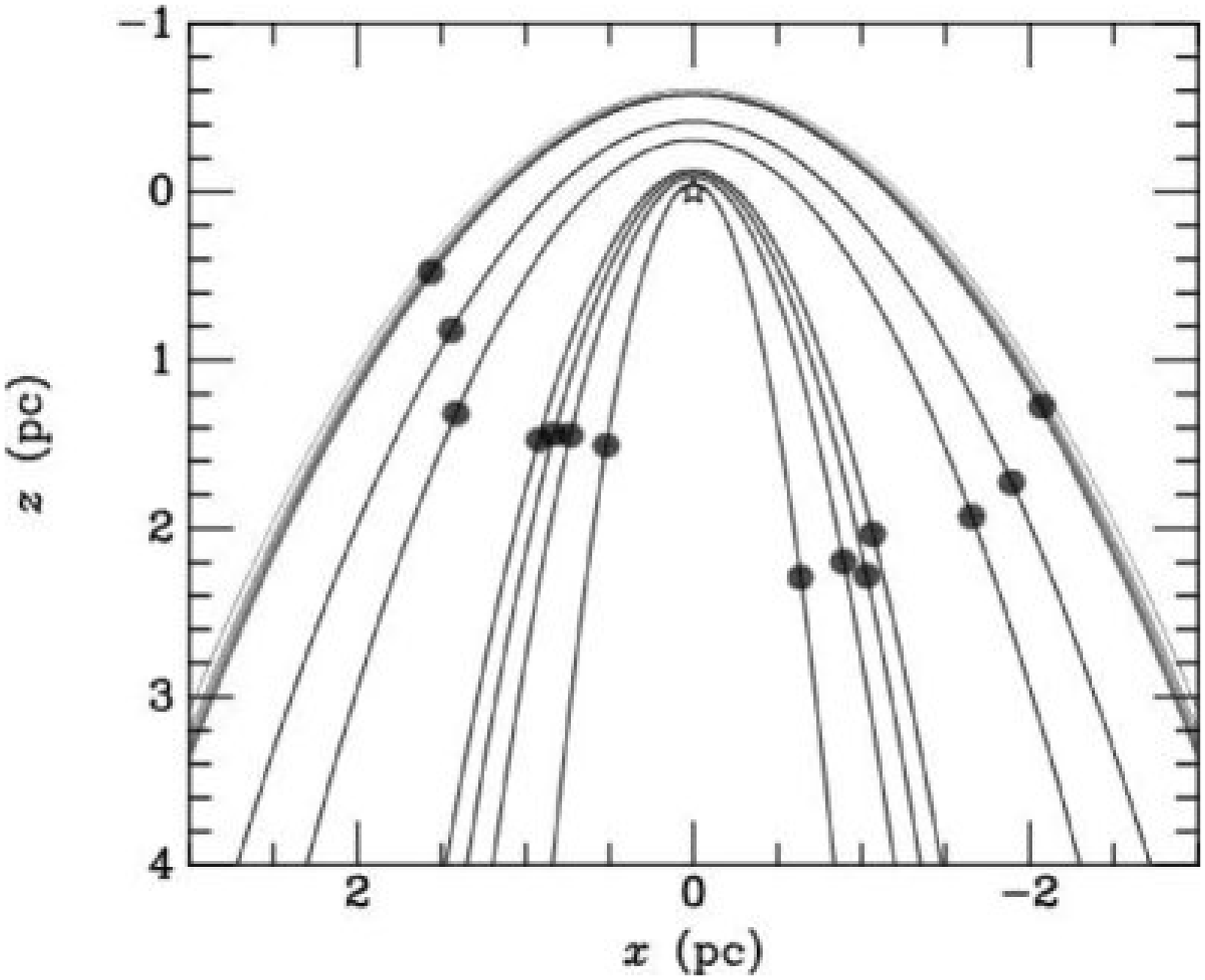} \hskip0.24in
\includegraphics[width=2.45in]{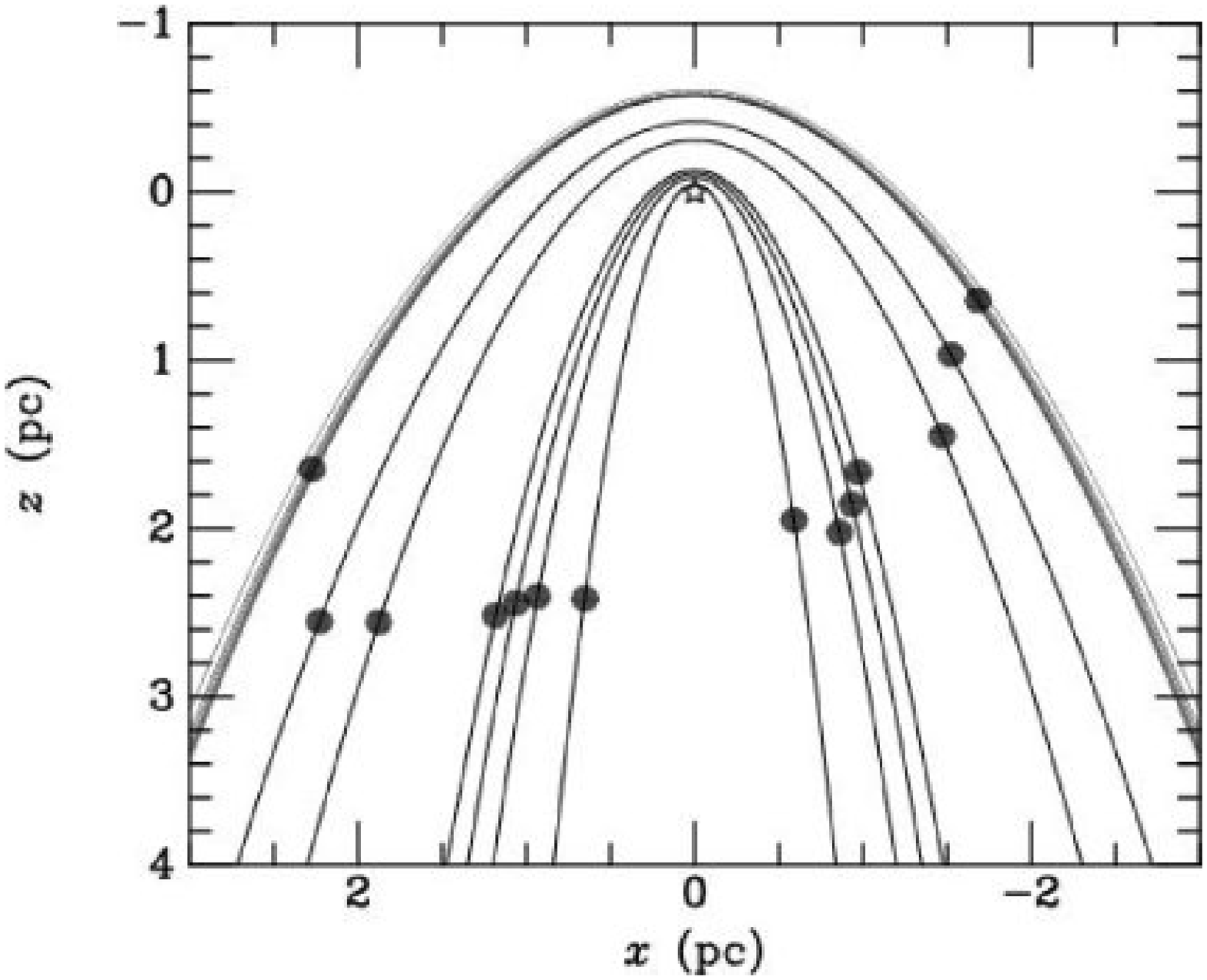}
\end{center}
\caption{Maps showing the locations of the outermost edges of the light echoes
seen in the available \HST\/ images from 2002 (inner parabolas), 2004 (two
intermediate parabolas), and 2005-6 (densely packed outer parabolas). The edges
are shown in a north-south plane centered on V838~Mon (left), and in an
east-west plane (right). Each parabola corresponds to the location of the
illuminated dust at that epoch, and the large black dots show the $(x,z)$
location of the outer edge along the corresponding parabola. Note that the $x$
and $z$ scales are in parsecs, and V838~Mon itself is located at the origin of
coordinates. Both maps suggest that the dust lies in an ellipsoidal
distribution, centered near the star.}
\end{figure}

Several authors concluded from the 2002 data that the dust edges defined a
plane, and suggested that this implied an interstellar origin for the dust.
However, with the inclusion of more recent observations, our Figures~3a and 3b
show that there is now clear curvature in the dust fronts. In fact, it appears
that the distribution is consistent with a roughly ellipsoidal structure, whose
center coincides approximately with the location of V838~Mon. Such a morphology
strongly suggests an origin from the star itself, rather than a feature that
arose by chance.

\item {\it Recent images appear to show an ``equatorial plane'' passing through
the star.} Figure~4a shows the \HST\/ image obtained on 2005 November 17. A
dashed line has been added to guide the eye to regions of enhanced surface
brightness that lie on both sides of the star. 

This image bears a remarkable similarity to that of a typical planetary nebula,
M27 or the ``Dumbbell Nebula.''  An amateur photograph of this nebula is shown
in Figure~4b, which clearly demonstrates an equatorial feature on either side
of the central star. 

\begin{figure}[htb]
\begin{center}
\includegraphics[height=2.4in]{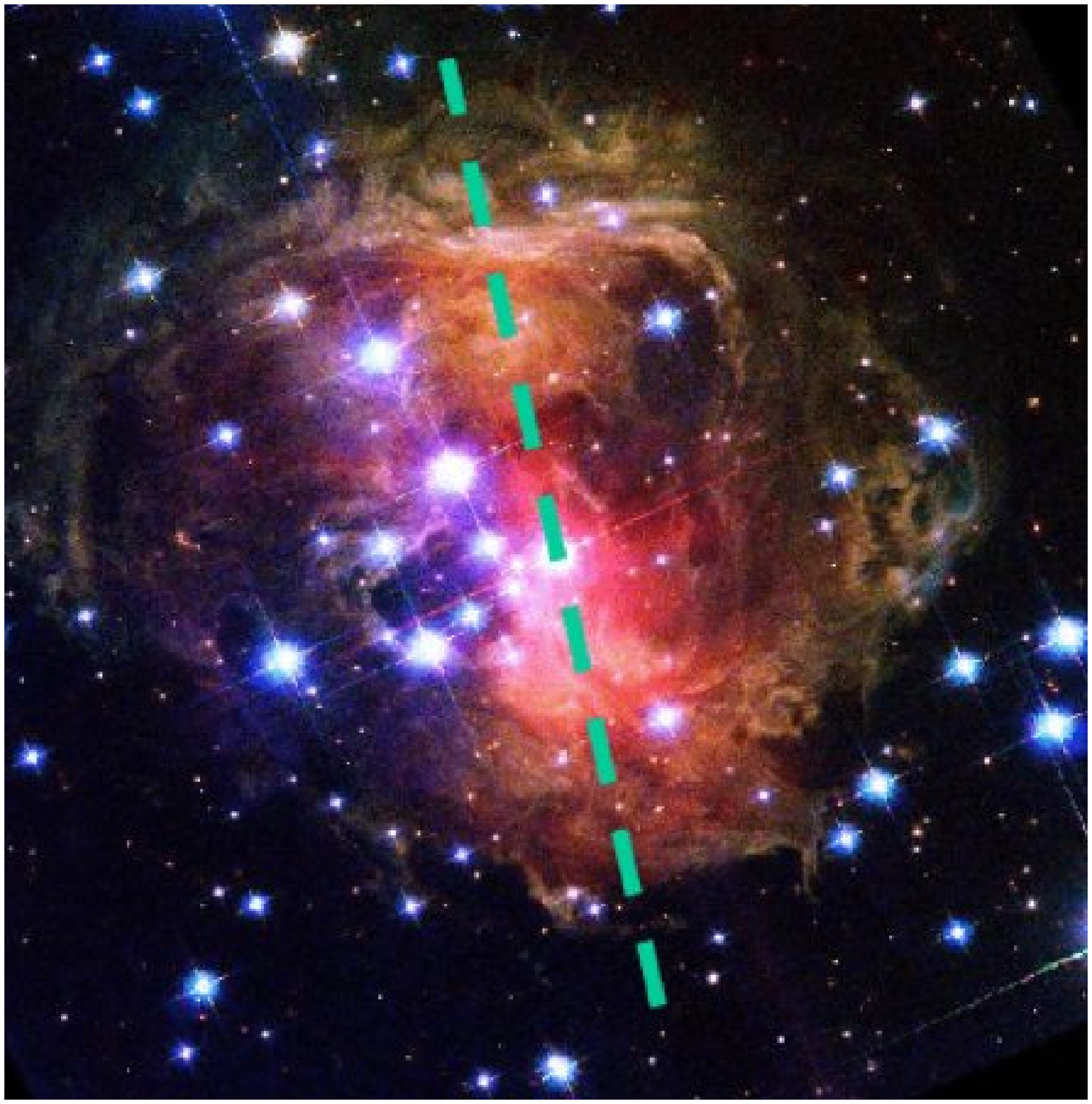} \hskip0.25in
\includegraphics[height=2.4in]{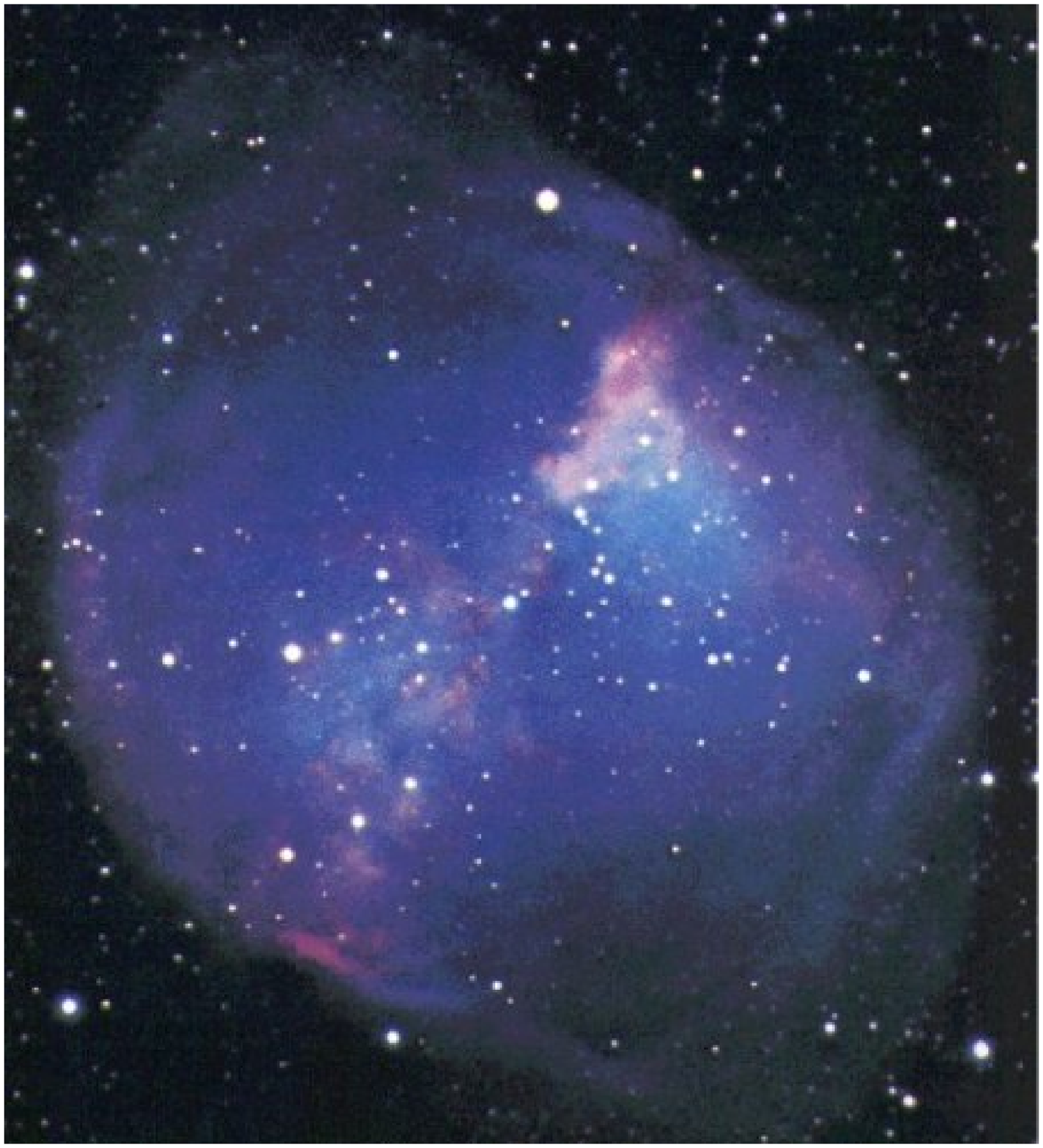}
\end{center}
\caption{Left: \HST\/ image of V838~Mon on 2005 November 17. The surface
brightness is highest along an ``equatorial plane'' passing through the star,
marked with a dashed line to guide the eye. Right: amateur photograph of the
planetary nebula M27. Note the striking resemblance of the light echo's
structure to that of the  planetary nebula.}
\end{figure}

There is no doubt that M27 was ejected from its central star. The striking
resemblance of the light echo in late 2005 to a genuine planetary nebula
provides another strong morphological argument that the illuminated dust was
indeed ejected from the star.

\end{enumerate}

\section{Conclusion}

V838 Monocerotis is illuminating the most spectacular light echoes seen in the
history of astronomy. The high spatial resolution provided by the {\it Hubble
Space Telescope\/} has yielded images of extraordinary beauty, as well as
providing unique scientific information.

The \HST\/ images have provided a direct geometrical distance to V838~Mon, based
on polarimetric imaging. Angular expansion rates have also given limits to the
distance. 

Several morphological features seen in the \HST\/ images strongly suggest that
the illuminated dust was ejected from the star in a previous outburst, similar
to the current one. The {\it Hubble\/} images show an extremely rich array of
filaments and other features on small physical scales, suggestive of
magnetohydrodynamic forces.

Future work on the \HST\/ images are expected to yield dust physics (i.e., the
angular and color dependence of the scattering function) and a fully
three-dimensional map of the dust distribution at high spatial resolution.

\acknowledgements 

I thank numerous colleagues for discussions and encouragement, including
especially the members of the {\it HST\/} V838~Mon observing team, and the
participants in this conference. Based on observations made with the NASA/ESA
{\it Hubble Space Telescope}, obtained at the Space Telescope Science Institute.
STScI is operated by the Association of Universities for Research in Astronomy,
Inc., under NASA contract NAS5-26555. Support from STScI grants including
GO-10913 is gratefully acknowledged.

\end{document}